\documentclass[useAMS,usenatbib,usegraphics, enumitem]{mn2e}
\usepackage{verbatim}
\usepackage{multirow}
\usepackage{graphicx,epsfig}  
\usepackage{amsmath,amssymb}
\usepackage{subfigure}
\usepackage{longtable}\usepackage{lscape}
\usepackage{times}
\newcommand{\integral}{{\it INTEGRAL}}
\newcommand{\rxte}{{\it RXTE}}

\newcommand{\swift}{{\it Swift}}
\newcommand{\sou}{{SwJ1745}}

\def\ergcms{erg cm$^{-2}$ s$^{-1}$ }
\def\ergs{erg s$^{-1}$}

\title[The failed outburst of SwJ1745]{Spectral and timing evolution of the bright failed outburst of the transient black hole Swift J174510.8--262411}

\author[M. Del Santo et al.]{
M. Del Santo$^{1}$\thanks{E-mail:melania@ifc.inaf.it}, T. M.  Belloni$^{2}$, J. A. Tomsick$^{3}$, B. Sbarufatti$^{4,2}$, M. Cadolle Bel$^{5}$, 
\newauthor
P. Casella$^{6}$, A. Castro-Tirado$^{7}$, S. Corbel$^{8}$, V. Grinberg$^{9}$, J. Homan$^{10,11}$, E. Kalemci$^{12}$, 
\newauthor
S. Motta$^{13}$, T. Mu\~noz-Darias$^{14,15}$,  K. Pottschmidt$^{16,17}$, J. Rodriguez$^{8}$, J. Wilms$^{18}$ 
\\
$^{1}$Istituto Nazionale di Astrofisica, IASF-Palermo, via Ugo La Malfa 153, Palermo, Italy \\
$^{2}$Istituto Nazionale di Astrofisica, Osservatorio Astronomico di Brera, Via E. Bianchi 46, I-23807, Merate, Italy\\
$^{3}$Space Sciences Laboratory, 7 Gauss Way, University of California, Berkeley, CA, 94720-7450, USA\\
$^{4}$Department of Astronomy and Astrophysics, Pennsylvania State University, University Park, PA, 16802, USA\\
$^{5}$Max Planck Computing and Data Facility, D-85748 Garching, Germany\\
$^{6}$Istituto Nazionale di Astrofisica, Osservatorio Astronomico di Roma, Via Frascati 33, 00040, Monteporzio Catone, Italy \\
$^{7}$Instituto de Astrofisica de Andaluc\'ia (IAA-CSIC), Glorieta de la Astronomia s/n, E-18008 Granada, Spain\\
$^{8}$Laboratoire AIM, UMR 7158, CEA/CNRS/Universit\'e Paris Diderot, CEA DSM/IRFU/SAp, F-91191 Gif-sur-Yvette, France\\
$^{9}$Massachusetts Institute of Technology, Kavli Institute for Astrophysics and Space Research, Cambridge, MA 02139, USA\\
$^{10}$MIT Kavli Institute for Astrophysics and Space Research, 77 Massachusetts Avenue 37-582D, Cambridge, MA 02139, USA\\
$^{11}$SRON, Netherlands Institute for Space Research, Sorbonnelaan 2, 3584 CA, Utrecht, The Netherlands\\
$^{12}$Faculty of Engineering and Natural Sciences, Sabanc\i\ University, Orhanl\i-Tuzla, 34956, Istanbul, Turkey\\
$^{13}$University of Oxford, Department of Physics, Astrophysics, Denys Wilkinson Building, Keble Road, Oxford OX1 3RH, UK\\
$^{14}$Instituto de Astrof\'isica de Canarias, 38205 La Laguna, Tenerife, Spain\\
$^{15}$Departamento de astrof\'isica, Univ. de La Laguna, E-38206 La Laguna, Tenerife, Spain\\
$^{16}$CRESST, University of Maryland Baltimore County, 1000 Hilltop Circle, Baltimore, MD 21250, USA\\
$^{17}$NASA Goddard Space Flight Center, Astrophysics Science Division, Code 661, Greenbelt, MD 20771, USA\\
$^{18}$Dr. Karl-Remeis-Sternwarte and Erlangen Centre for Astroparticle Physics, Friedrich Alexander Universit\"{a}t Erlangen-N\"{u}rnberg,\\
Sternwartstr. 7, 96049 Bamberg, Germany
}

\begin{document}
\date{Accepted 2015 December 7.  Received 2015 December 7; in original form 2015 September 22.}

\pagerange{\pageref{firstpage}--\pageref{lastpage}} \pubyear{2015}

\maketitle

\label{firstpage}

\begin{abstract}
\indent
We studied time variability and spectral evolution of the Galactic black hole transient Swift J174510.8--262411
during the first phase of its outburst. \integral\ and \swift\ observations collected from 2012 September 16 until October 30 have been used.
The total squared fractional rms values did not drop below 5\% and QPOs, when present, were type-C,
indicating that the source never made the transition to the soft-intermediate state.
Even though the source was very bright (up to 1 Crab in hard X-rays), it showed a so called failed outburst as it never reached the soft state.
XRT and IBIS broad band spectra, well represented by a hybrid thermal/non-thermal Comptonisation model, 
showed physical parameters characteristic of the hard and intermediate states. In particular, the derived temperature of the geometrically
 thin disc black body was about 0.6 keV at maximum. 
We found a clear decline of the optical depth of the corona electrons (close to values of 0.1), as well as of the total compactness ratio $\ell_{\rm h}/\ell_{\rm s}$.
The hard-to-hard/intermediate state spectral transition is mainly driven by the increase in the soft photon flux in the corona, rather than small variations of the electron heating.
This, associated with the increasing of the disc temperature,  is consistent with a disc moving towards the compact object scenario, i.e. the truncated-disc model.
 Moreover, this scenario is consistent with the decreasing fractional squared rms and increasing of the noise and QPO frequency.
In our final group of observations, we found that the contribution from the non-thermal Comptonisation to the total power supplied to the plasma 
is $0.59^{+0.02}_{-0.05}$  and that the thermal electrons cool to kT$_{e} < 26 $ keV. 
\end{abstract}


\section{Introduction}\label{sec:intro}
Galactic black-hole binaries (BHB) emit strong X/$\gamma$-ray radiation when accreting matter from the stellar companion.
Most of them are transients, i.e. they  spend most of the time in a dim, quiescent state, displaying
X-ray luminosities as low as L$_{\rm X} \sim 10^{31}$ \ergs, spaced out by episodic outbursts during which the sources show X-ray luminosities of  L$_{\rm X} \sim 10^{36-39}$ \ergs.
Based on the different X/$\gamma$-ray spectral properties, BHBs  are known to show different
spectral states over their outbursts \citep{zg04, mcclintock06}. 
Usually, the spectral variability is interpreted as being due to changes in the geometry of  the central parts of the accretion flow \citep{zdz00, done07}.

At the beginning of the outburst, these sources are in the hard state (HS) with the spectrum roughly described 
by a dominant cut-off power-law (typically, photon index $\Gamma$ $\sim$1.5 and high energy cut-off $E_{\rm cut} \sim$100 keV) 
and (often) a faint soft thermal component with a black-body temperature kT$_{\rm in} < $  0.3 keV ascribed to the emission from an
accretion disk truncated at large distances from the central BH ($\sim$100 km; \citealt{done07}).
The hard X-ray emission in the HS is believed to originate from thermal Comptonisation of soft disc-photons 
in a hot electron cloud \citep{eardley75, sunyaev80}. 
Thanks to the large area of the \rxte\ satellite, it was observed that the spectral states are related to the timing properties  \citep{homan05,belloni11}.
The power density spectrum (PDS) of sources in the hard state can be decomposed into a number of broad 
Lorentzian components and sometime with type-C quasi periodic oscillations (QPO).
The type-C QPOs are characterised by a strong (up to 16\% rms), narrow ($\nu/\Delta \nu \sim$7--12), and variable peak frequency. A subharmonic 
and a second harmonic peak are sometime seen \citep{casella05, wij99}. 
Strong band-limited noise components with rms values of about  30\% are observed 
 and the radio emission indicates the presence of  a steady compact jet \citep{fender01}.\\
\begin{figure}
\centering
\includegraphics[height=13cm,width=8cm]{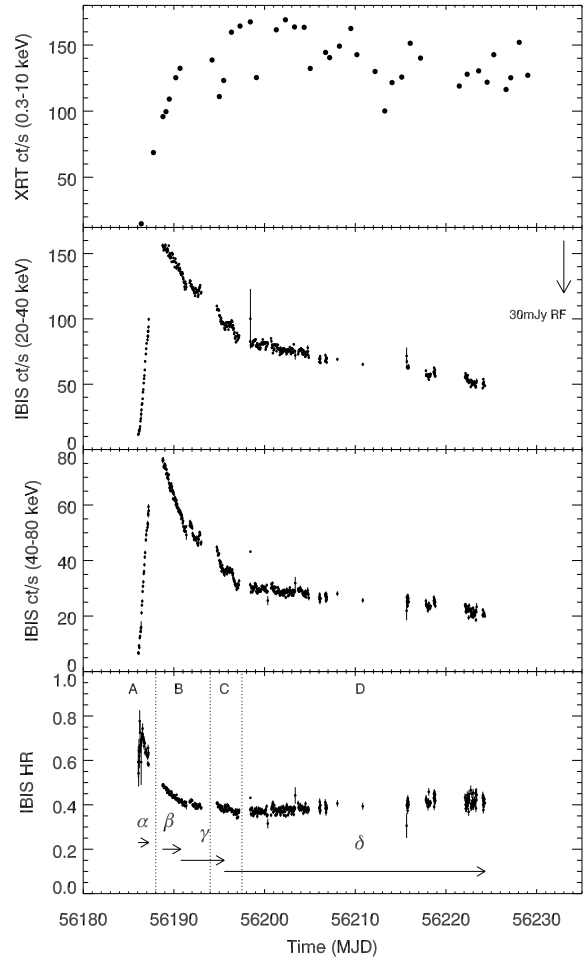}
\caption{From the top: XRT count rate extracted after the pile-up correction in the energy range 0.3--10 keV (errors are smaller than the symbols size);
IBIS/ISGRI count rate in the energy range 20-40 keV and 40-80 keV; IBIS/ISGRI hardness ratio defined as the ratio of  the hard rate divided by the soft one. 
In the panel 2 (from the top), the time of the radio flares reported in \citet{curran14} is marked. In the IBIS HR panel,  the periods used to the joint XRT--IBIS  (A, B, C, D)
and JEM-X--IBIS ($\alpha, \beta, \gamma, \delta$) spectral analysis are indicated.
 \label{fig:lc}}
\end{figure}
As the outburst progresses, the X-ray and radio luminosities both increase, but the X-ray colour of the spectrum remains hard \citep{corbel05}.
 Then, the transient BHB evolves into an intermediate state  at an  almost constant flux level
with spectral parameters in between the two main states. The disc black-body shows typical temperatures of kT$_{\rm in} \simeq $  0.3-0.5 keV,
while the hard X-ray spectra are usually explained with the hybrid thermal/non-thermal Comptonisation (e.g. \citealt{zdz04, done07, delsanto08}). 
Based on quite complex timing characteristics, \citet{homan05} identified two different intermediate states, namely the HIMS and the soft-intermediate state (SIMS).
Despite slight spectral softening, the PDS before and after the HIMS-to-SIMS transition showed significant differences:
from rms values about 10-20\% and strong type-C QPOs to much weaker noise (a few \%) and (possibly) type-B QPOs
(see \citealt{belloni11} for a recent review). 
However, based on the X/$\gamma$-ray spectra of BHBs, a firm separation between HIMS and SIMS cannot be established
since this transition can display different properties in different outbursts \citep{delsanto08, delsanto09}.\\
Thereafter, most BHBs reach the second main state, namely soft state (SS), which is  characterised by a thermal component, i.e. the multi-color black-body emission  
from an accretion disc  \citep{ss73} 
peaking at a few keV and much softer high energy power law ($\Gamma > 2.5$) \citep{zdz00}.
In the framework of the truncated-disc model, in the SS, the geometrically thin accretion disc is believed to extend down to the innermost stable circular orbit
(ISCO)\footnote{Note however that a number of papers report on discs remaining at the ISCO in hard state \citep{miller06, reis10, reynolds10, miller12}.} 
and it is the source of soft seed photons for non-thermal Comptonisation in small active coronal regions located above and below the disc \citep{zg04}. 
In this state,  the variability is in the form of a weak (down to 1\% fractional rms) steep component. Very weak QPOs are sometimes detected in the
10-30 Hz range \citep{belloni10}.
When the flux starts to decrease a reverse transition occurs until the source is back to the HS then to quiescence.

\begin{table*}
\begin{center}
\caption{Observations log of the 26 IBIS data-set used for the spectral modelling. Group name (Group), \integral\ revolution (Rev), total number of pointings ($N_{SCW}$), 
grouping step  of the pointings (Step), effective observing time (Obs), number of spectra obtained in the related group (Sub-group), start time and end time of the 4 groups are shown.
($^{*}$)The last spectrum, D4, has been obtained by averaging the last 104 SCWs.}\label{tab:log}
\begin{tabular}{lccclccc} 
\hline
Group & Rev  & $N_{SCW}$  & Step &  Obs & Sub-group  &  Start  & End \\
         &            &              &          &    [ks]    &     &  &      \\
\hline

A &  1212     &       30   &   5    &     $\simeq  57$    &    6  &   2012-09-16T01:15:58   & 2012-09-17T05:34:09  \\
B & 1213-1214   & 100 & 10  & $\simeq 72 $ &       10   &  2012-09-18T17:34:10    &   2012-09-23T01:02:18  \\
C & 1215            &     60  & 10 &    $\simeq 43 $  &    6    &   2012-09-24T17:07:31   &     2012-09-27T04:59:38  \\
D & 1216-1224   &   248    &    48($^{*}$)  & $\simeq 467$ &   4      &   2012-09-28T09:28:38   &     2012-10-24T07:38:54       \\

\hline
\end{tabular}
\end{center}
\end{table*}

Despite the general pattern followed during most of the outbursts of transient BH binaries,
a number of unusual outbursts, in which sources do not show soft states, have been observed.
There are different types of these so called "failed" outbursts:
a number of sources never leave the HS during the outburst \citep{brock04},
while others proceed to an intermediate state before returning to the hard state and quiescence \citep{capitanio09, ferrigno12, soleri13}.
It is worth noting that some sources have undergone both canonical outbursts and failed outbursts \citep{sturner05}.
Since most of the failed outburst are under-luminous, 
the lack of soft-state transitions is possibly connected to a premature
decrease of the mass accretion rate, as during the 2008 outburst of H1743--322 \citep{capitanio09}.

The X-ray source Swift J174510.8--262411 (hereafter \sou) was discovered by the Burst Alert Telescope (BAT; \citealt{barthelmy05}) 
on board the \swift\ satellite on 2012 September 16 \citep{cummings12}.
Almost simultaneously observed by the \integral\ satellite, \sou\ appeared immediately as a bright Galactic BHB,
 since the 20--40 keV flux increased from 63 mCrab up to 617 mCrab  in one day \citep{vovk12}.
Soon after a number of multi-wavelength  campaigns, including \swift/XRT \citep{tom12}, were performed. 
Optical spectroscopy and photometry inferred an orbital period $\lesssim$21 h, a companion star with a spectral type later than A0 and a
distance closer than $\sim$ 7 kpc \citep{munoz13}.
Multi-frequency data from radio arrays showed that a discrete ejection event occurred for the first time 
in a "failed" outburst (\citealt{curran14}; see radio flares indicated in Fig. \ref{fig:lc}).
These events are expected in the intermediate state when the source crosses the 'jet-line'  (\citealt{fender09} and ref. therein).
As reported in \citet{fender09}, the time coincidence of the crossing of the jet line and the HIMS-SIMS transition is not exact.
However, in \sou\ the SIMS seems to be never reached.

In this paper we present spectral and timing results of two Target of Opportunity campaigns performed during the first part of the \sou\ outburst, 
i.e. 1 Ms with \integral\ (P.I.: T. Belloni) and a number of \swift/XRT pointings (P.I.s: M. Del Santo and B. Sbaruffatti),
while the decay of the same outburst (beginning of 2013) is reported in \citet{kalemci14}.

\begin{table}
\begin{center}
\caption{Observations log of the JEM-X2 data-set.}
\label{tab:logj}
\begin{tabular}{lccc} 
\hline
Group & Rev  & $N_{SCW}$  &  Obs \\
         &            &                  &    [ks]      \\
\hline
$\alpha$ &  1212     &      5   &      $\simeq  16 $    \\
$\beta$ & 1213     & 13  &      $\simeq 43 $  \\
$\gamma$ & 1213-1215  &     17 &    $\simeq 56 $  \\
$\delta$ & 1215-1224   &   81    &     $\simeq 217$  \\
\hline
\end{tabular}
\end{center}
\end{table}

\section{Observations and data reduction}\label{sec:data}

\subsection{INTEGRAL}

We have analysed the \integral\ \citep{win03} data of \sou\ collected in the period 2012 September-October.
The IBIS \citep{ube03} data-set  has been obtained selecting all observations including 
\sou\ in a partially coded field of view (FOV) of $15^{\circ}  \times 15^{\circ}$ where the instrument response is well known.
This resulted in 438 pointings (Science Window, SCW) from \integral\ revolution 1212 up to 1224.
The \integral\ data analysis and reduction has been performed with the off-line analysis software, OSA v.10.1 \citep{goldwurm03,courvoisier03}.
After the data reduction and correction (i.e. dead time), the total IBIS/ISGRI \citep{lebrun03} effective observing time is roughly 650 ks.
The IBIS/ISGRI light curve in the 20--40 keV and 40--80 keV bands and the related hardness ratio are shown in Fig. \ref{fig:lc}.
Spectra by SCW (1.7--3.5 ks of duration) have been extracted in 62 channels from 20 keV up to 500 keV.

\begin{table*}
\caption{\swift/XRT log table of the pointings used in this work. The target ID is 533836. The average rate reported in this table has been extracted before the pile-up correction.
 ($^{*}$)These pointings have been used in the timing analysis only (sequence \#018 is affected by a larger than usual pointing error, while the \#045 shows a too low statistics to perform a spectral fitting).}
\label{tab:xrt_journal}
\begin{tabular}{llllll}
\hline
Seq. \# &  Beginning of obs.         & End of obs.     & Exposure (s)  & Average Rate (cps) & Inner extraction radius (pix)            \\
\hline      	
  000 &    2012-09-16T09:37:11  &  2012-09-16T14:25:47  &  6502  & 18  &   0   \\ 	 
  002 &    2012-09-17T17:49:22  &  2012-09-17T19:32:17  &  996  & 93  &   1  \\
  003 &    2012-09-18T19:03:30  &  2012-09-18T22:18:50  &  990 & 131 &   1   \\ 
  005 &    2012-09-19T11:43:28  &  2012-09-19T16:21:57  &  982  & 152 &   1   \\ 
  006 &    2012-09-19T03:05:21  &  2012-09-19T05:18:15  &  976  & 158 &   1   \\ 
  007 &    2012-09-20T04:55:43  &  2012-09-20T05:09:58  &  833  & 171 &   1   \\ 
  008 &    2012-09-20T16:17:29  &  2012-09-20T16:34:58  &  1036 & 180 &   1   \\ 
  009 &    2012-09-24T04:59:09  &  2012-09-24T05:07:09  &  470  & 241 &   2   \\ 
  011 &    2012-09-25T00:12:25  &  2012-09-25T00:30:58  &  1096  & 268 &   2   \\ 
  012 &    2012-09-25T11:49:39  &  2012-09-25T12:05:58  &  966  & 312 &   3   \\ 
  013 &    2012-09-26T08:16:11  &  2012-09-26T08:36:58  &  1221  & 297 &   2   \\ 
  014 &    2012-09-27T07:02:45  &  2012-09-27T07:21:58  &  1140  & 359 &   3   \\ 
  015 &    2012-09-28T10:11:58  &  2012-09-28T10:16:58  &  278  & 363 &   3   \\ 
  016 &    2012-09-29T02:22:06  &  2012-09-29T02:38:58  &  993  & 352 &   3   \\ 
  017 &    2012-10-01T06:58:20  &  2012-10-01T07:14:58  &  969  & 384 &   3   \\ 
 018$^{*}$ &    2012-09-30T08:39:21  &  2012-09-30T08:56:58  &  103  & 477 &   4   \\ 
  019 &    2012-10-02T07:09:55  &  2012-10-02T07:28:58  &  1118  & 385 &   3   \\ 
  020 &    2012-10-03T07:14:03  &  2012-10-03T07:30:58  &  992  & 352 &   3   \\ 
  021 &    2012-10-04T08:45:34  &  2012-10-04T09:04:58  &  1143  & 360 &   3   \\ 
  022 &    2012-10-05T00:48:34  &  2012-10-05T01:03:58  &  901  & 398 &   3   \\ 
  023 &    2012-10-07T04:07:27  &  2012-10-07T04:21:57  &  857  & 412 &   4   \\ 
  024 &    2012-10-06T17:22:57  &  2012-10-06T18:34:21  &  972  & 390 &   3   \\ 
  025 &    2012-10-08T06:01:15  &  2012-10-08T06:17:55  &  973  & 367 &   3   \\ 
  026 &    2012-10-09T12:16:24  &  2012-10-09T12:31:58  &  913  & 360 &   3   \\ 
  027 &    2012-10-10T04:16:42  &  2012-10-10T04:32:58  &  968  & 353 &   3   \\ 
  029 &    2012-10-12T04:24:09  &  2012-10-12T04:39:58  &  939  & 311 &   3   \\ 
  030 &    2012-10-13T06:02:46  &  2012-10-13T06:19:58  &  1020  & 294 &   2   \\ 
  031 &    2012-10-14T01:17:47  &  2012-10-14T01:33:58  &  961  & 227 &   2   \\ 
  032 &    2012-10-15T02:57:15  &  2012-10-15T03:12:58  &  922  & 295 &   3   \\ 
  033 &    2012-10-16T01:24:17  &  2012-10-16T01:39:58  &  932  & 271 &   2   \\ 
  034 &    2012-10-17T04:39:41  &  2012-10-17T04:55:58  &  962  & 280 &   2   \\ 
  035 &    2012-10-22T08:04:47  &  2012-09-17T19:32:17  &  964  & 226 &   2   \\
  036 &    2012-10-21T11:14:04  &  2012-10-21T11:32:58  &  1110  & 292 &   2   \\ 
  037 &    2012-10-23T14:32:37  &  2012-10-23T14:47:57  &  915  & 268 &   2   \\ 
  038 &    2012-10-24T12:59:51  &  2012-10-24T13:14:58  &  899  & 319 &   3   \\ 
  039 &    2012-10-25T06:38:09  &  2012-10-25T06:53:58  &  923 & 276 &   2   \\ 
  040 &    2012-10-26T15:04:12  &  2012-10-26T15:20:57  &  980  & 339 &   3   \\ 
  041 &    2012-10-27T03:31:17  &  2012-10-27T03:46:58  &  920  & 301 &   2   \\ 
  042 &    2012-10-22T08:20:58  &  2012-10-28T02:13:58  &  983   & 291 &   2   \\
  043 &    2012-10-29T00:28:40  &  2012-10-29T00:42:58  &  845  & 269 &   2   \\ 
  045$^{*}$  &  2012-10-31T13:38:44  &  2012-10-31T13:55:58  &  1029  & 360 &   3  \\
\hline                                                                                          
\hline                                                                                          
\end{tabular}                                                                                   
\end{table*}

\begin{figure}
\centering
\includegraphics[height=8cm]{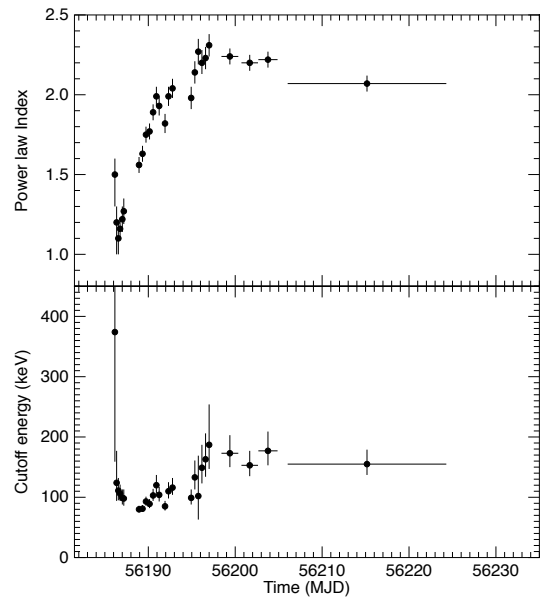}
\caption{Power-law slope and high energy cut-off evolution of the 26 IBIS/ISGRI spectra.
 \label{fig:cutpl}}
\end{figure}

\begin{figure}
\centering
\includegraphics[height=10cm]{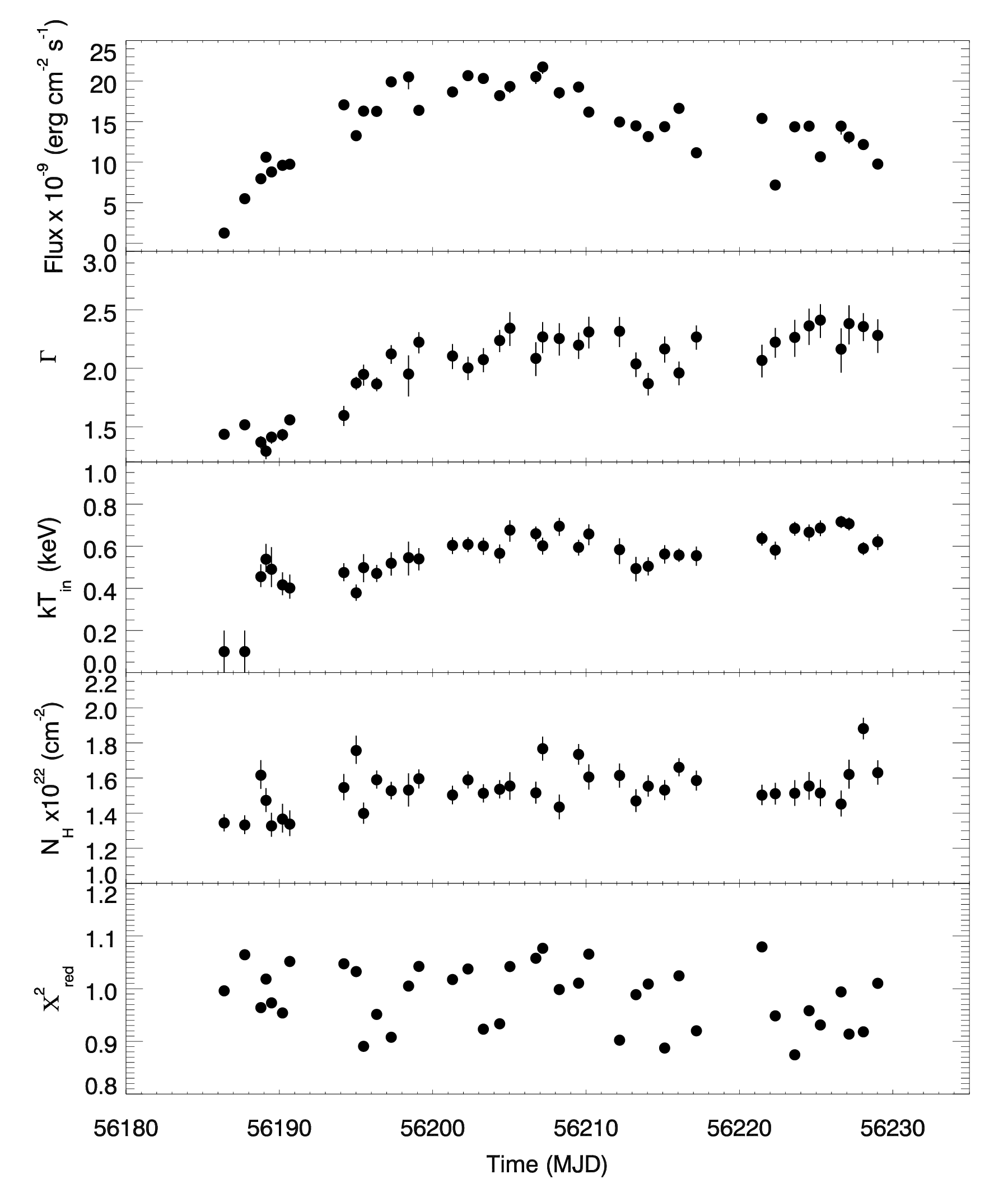}
\caption{XRT parameters of the 39 spectra fitted with an absorbed power law plus a disc black-body. 
From top to bottom: 2-10 keV absorbed flux, power-law slope, disc temperature, column density and reduced $\chi^{2}$.
 \label{fig:po}}
\end{figure}

In order to increase the statistics, the 438 spectra were averaged as follows.
First, based on the discontinuities in the observations we identified 3 groups (A, B and C, see Fig. \ref{fig:lc}, bottom);
the last group (D)  contains all observations (even those that were not continuous) of the last part of the outburst observed by \integral\  in 2012 (Fig. \ref{fig:lc}, bottom).
Then, within each group, spectra have been averaged as reported in Tab. \ref{tab:log} resulting in 26 final spectra. 
Because of the low statistics  the last sub-group, i.e. D4, includes 104 spectra instead of 48.
Due to the known IBIS/ISGRI calibration issue related to the energy reconstruction, 
systematics of about 3-4\% are required by the spectral fits to avoid large residuals at energy around 50 keV and 100 keV.

Quasi-simultaneous \integral/JEM-X \citep{lund03} data have been analysed. 
Because of the smaller JEM-X FOV, the total number of analysed pointings is 116 (see Tab. \ref{tab:logj}).
JEM-X2 spectra have been extracted in 32 channels.
Although recommendation from the JEM-X calibration team is to use the energy range 5-22 keV,
we fitted spectra from 6 keV up to 18 keV by adding systematics of 3\% 
because of additional residuals appearing especially in bright and variable sources.

\subsection{Swift}
The \swift/XRT monitoring campaign of the outburst of \sou\ was performed mainly in Windowed Timing (WT) observing mode. 
The XRT observations used for this paper were performed quasi-simultaneously with the \integral\ ToO campaign,
i.e. from September 16 to October 30, 2012 (see XRT count rate in Fig. \ref{fig:lc}).  A journal of the observations is given in Table  \ref{tab:xrt_journal}.\\
To obtain XRT spectra, data were processed using the FTOOLS software package distributed
inside HEASOFT (v6.16)  and the related calibration files from the
NASA Calibration Database. We ran the task \textit{xrtpipeline} applying 
calibrations and standard filtering criteria.
Events with grades 0 were selected, in order
to reduce the effect of energy redistribution at low energies that is
known to affect XRT data for bright, heavily absorbed sources\footnote{see http://www.swift.ac.uk/analysis/xrt/digest\_cal.php for details}.
For the same reason spectral analysis was performed in the 1--10 keV energy band. Due to the high count rates, the data are
affected by pile-up starting with sequence 1. The pile-up correction for the WT data was performed following the same procedure
used by \citet{romano06}, determining the size of the region to be excluded from the photon extraction studying the event grade distribution. 
Source photons were thus extracted
from an annular region with an outer radius 30 pixels (1 pixel = 2''.36)
and an inner radius 0 to 4 pixel, depending on count rate (see Tab.  \ref{tab:xrt_journal}).

According to the XRT calibration document\footnote{see http://heasarc.gsfc.nasa.gov/docs/heasarc/caldb/swift/docs/xrt/SWIFT-XRT-CALDB-09\_v19.pdf},
high signal-to-noise WT spectra typically show
residuals of about 3\% near  the  gold edge (2.2 keV) and the silicon edge (1.84 keV),
and can be as high  as 10\%. We obtained good fits by including 3\% of systematics in the XRT spectra and in a few cases (when the residuals around the edges were 
higher) we ignored the channels between 1.8 and 2.4 keV.

From each XRT observation we have extracted a light curve with 0.003532 s 
time resolution from a region 40 pixels wide centred on the source. For timing analysis, we produced Leahy-normalized Power Density Spectra (PDS) from 128 s segments and averaged them, 
obtaining one average PDS per observation. The power spectra cover the frequency range from 7.8125 mHz to 142 Hz. 
We subtracted the Poissonian noise contribution estimated as the mean power level above 20 Hz, where no source signal was seen. 
We converted the PDS to squared fractional rms \citep{belloni90}.

\begin{figure*}
\includegraphics[height=16cm, width=10cm]{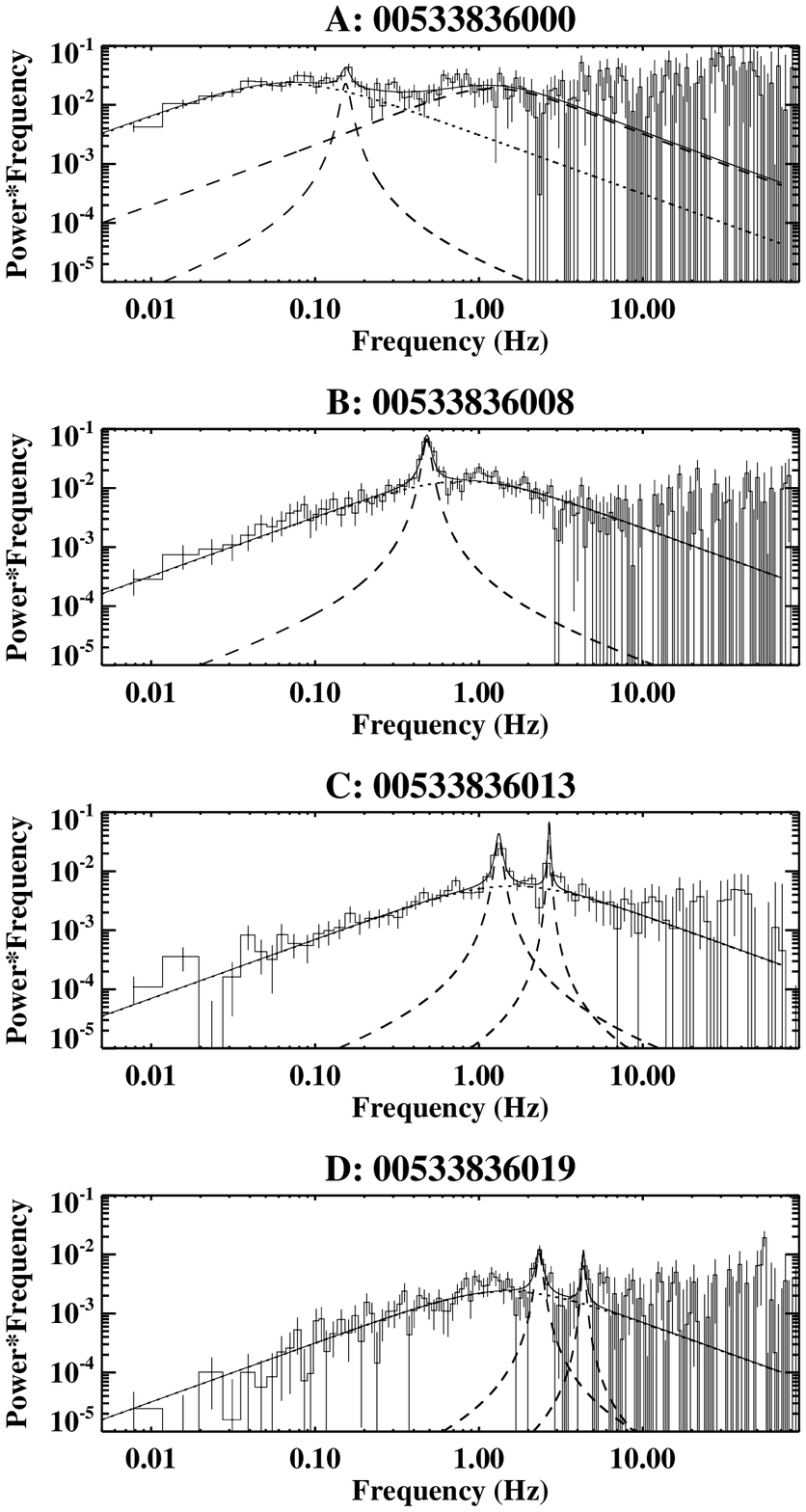}
\hspace{+1cm}
\includegraphics[height=16cm,  width=6cm]{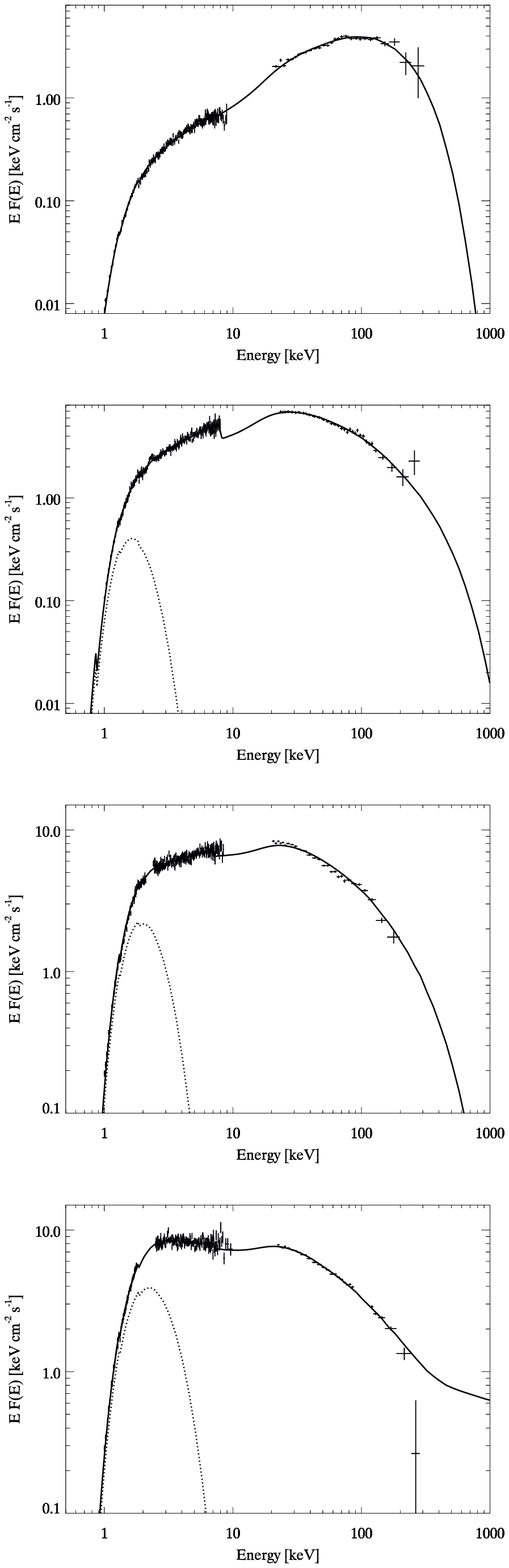}
\caption{{\it Left}: XRT power density spectra of four pointings, i.e. \#000, \#008, \#013, \#019
fitted with a zero-centered Lorentzian plus one (or two) QPOs.
{\it Right}: Joint ISGRI-XRT energy spectra of four different sub-groups, i.e. A3-000, B5-008, C5-013, D2-019, fitted with the {\it eqpair} model (solid line).
The curve has been normalised taking into account an IBIS cross-normalisation factor roughly ranging between 0.6 and 1.
The disk black-body component is also shown (dotted line).
Spectral parameters are reported in Tab. \ref{tab:fit}. 
\label{fig:spec_time}}
\end{figure*}

\begin{figure*}
\centering
\includegraphics[height=6cm]{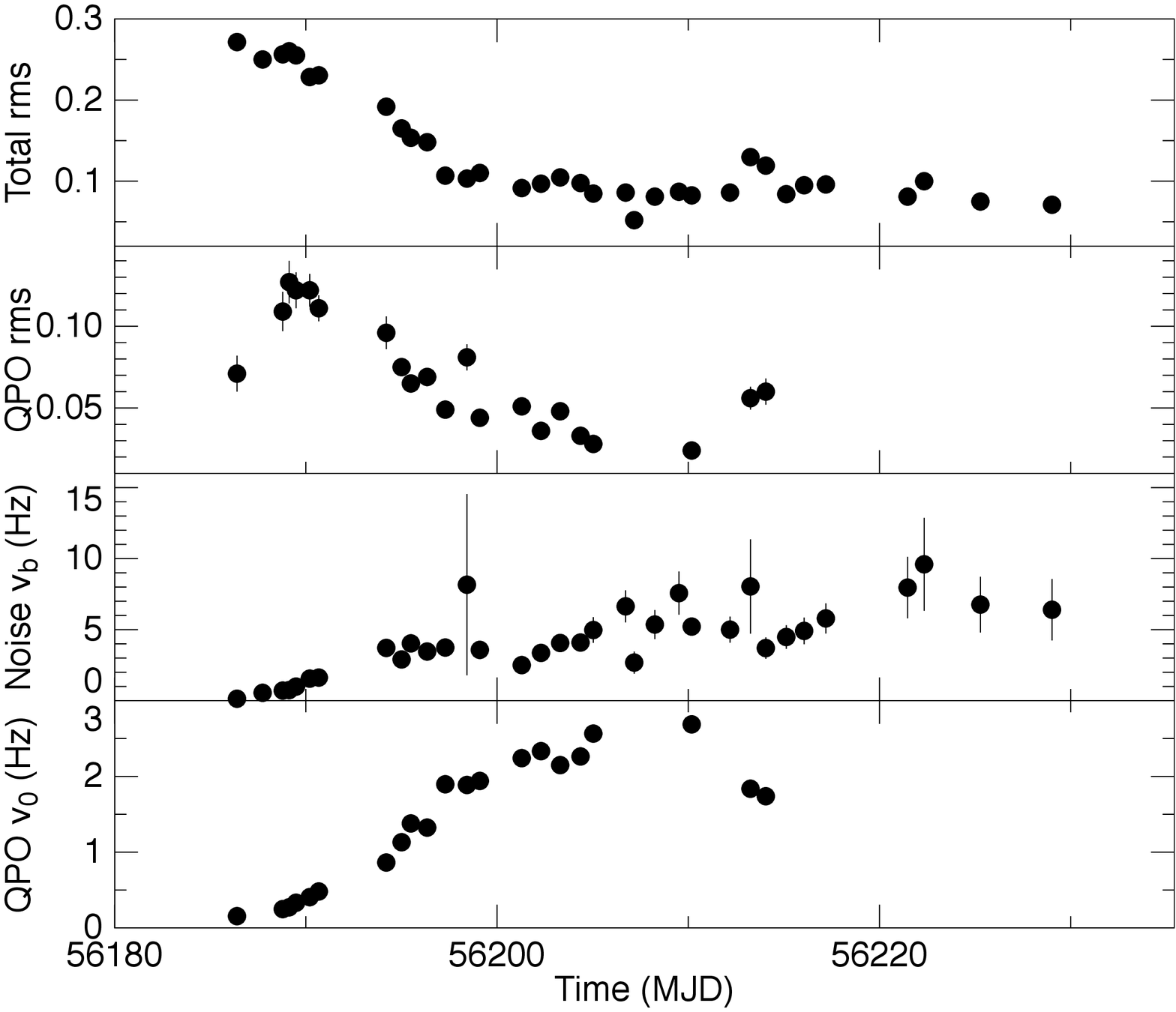}
\includegraphics[height=6cm]{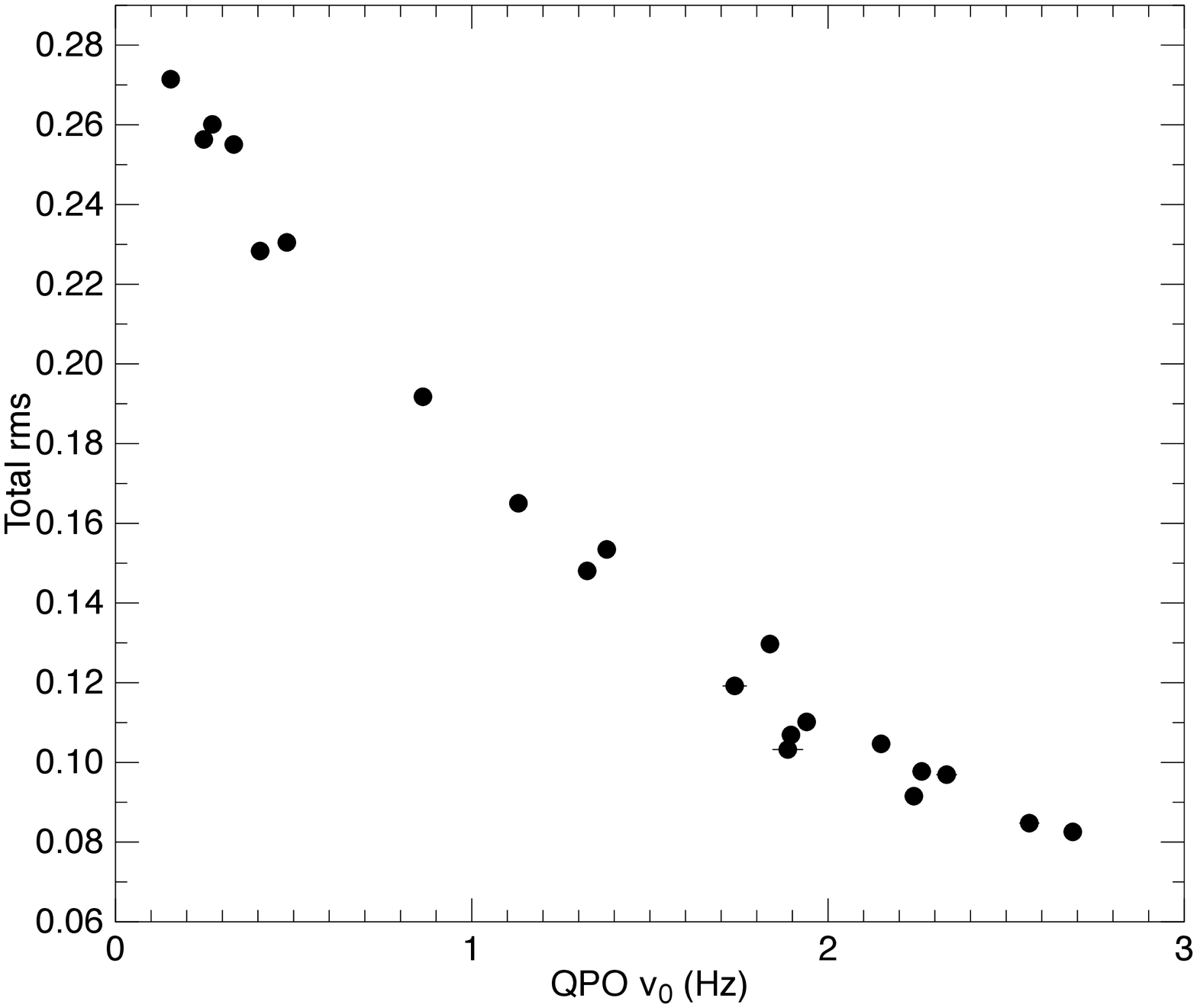}
\caption{{\it Left}: evolution of the main timing parameters (total rms uncertainties smaller than symbol size). {\it Right}: correlation between QPO frequency and the total rms.
 \label{fig:timing}}
\end{figure*}

\section{Spectral analysis with simple models}
First of all, we  fit separately the 26 IBIS/ISGRI spectra and the 39 XRT spectra by using simple 
models, such as cutoff power-law (Fig. \ref{fig:cutpl}) and an absorbed ({\it tbabs} in {\sc xspec}) 
multicolor disc black-body ({\it diskbb} in {\sc xspec}) plus a power law (Fig. \ref{fig:po}), respectively.
We used cross sections of \citet{verner96} and abundances of \citet{wilms00} for the interstellar absorption
and left N$_{H}$ as a free parameter.

The power-law of the IBIS/ISGRI hard X-ray spectra (Fig. \ref{fig:cutpl}) steepens as the cut-off energy decreases  in group A, until MJD 56188.
Then, starting from group B,
the cutoff starts to increase again until it stalls.
This behaviour of the cut-off during the BHB outbursts has been already observed in GX 339-4 by \citet{motta09}.
However, cut-off power-law ({\it cutoffpl} in {\sc xspec}) is an empirical model, which is only roughly related to the electron temperature of the Comptonising plasma.
In complex spectra, i.e. when additional components play a role (non-thermal Comptonisation, reflection),
the high energy cut-off does not reflect the evolution of the electron temperature of the corona. 
Indeed, in Section \ref{sec:eqpair}, we present results by using a physical model, i.e. the hybrid Comptonisation model {\it eqpair}.\\

In Fig. \ref{fig:po}, we show the evolution of the spectral parameters of the soft X-ray spectra.
While the disc black-body component is not needed in the first two spectra, 
its addition improves considerably the spectral fit of the following spectra (from MJD 56188). 
In spectrum \#003 the F-test probability is 6.5$\times 10^{-22}$.
The inner disc black-body temperature (kT$_{\rm in}$) varies between 0.4 and 0.7 keV, the power-law photon index steepens from 1.5 up to 2.2 and then it stays constant,
the N$_{\rm H}$ varies roughly between 1.4 and 1.7 during this part of the outburst.
We note that the value of this parameter is significantly different than the one found by \citet{kalemci14} in the hard state during the outburst decay ($\sim$2.2$\times10^{22}$ cm$^{-2}$). \\

Based on the XRT and ISGRI spectral behaviour, we conclude that the HS-to-HIMS spectral transition occurred during the gap (between A and B) in the  \integral\ observations
and since XRT pointing \#003 (roughly at MJD 56188).
This is also confirmed by the timing analysis (see Sec. \ref{sec:timing}).

\section{Timing analysis results}
\label{sec:timing}
Each PDS was fit with a zero-centered Lorentzian (see \citealt{belloni02}) and inspected for residuals. 
Then, PDSs were fitted with a zero-centered Lorentzian plus a QPO, also modeled as a Lorentzian (Fig. \ref{fig:spec_time}B). 
The starting frequency was based on the residuals, and the starting width was typically 0.05-0.1 Hz (the final best fit parameters do have some dependence on the starting parameters). 
In some cases, the QPO is not statistically significant ($< 3 \sigma$). If another noise peak (broad or narrow) appeared in the residuals,  
then we fit the PDS with a combination of three Lorenztians, one of which zero-centered (Fig. \ref{fig:spec_time}C and \ref{fig:spec_time}D).
This procedure led to the detection of a single QPO peak in 20 observations and two QPO peaks in about ten observations, 
where the second peak is consistent with being the first overtone of the first peak. 
However, only in a few observations the second QPO seems to be significant ($ \geq 3 \sigma$).
  A broad component in addition to the zero-centered Lorentzian plus a faint QPO is observed in the HS, pointing  \#000 (see Fig. \ref{fig:spec_time}A).
  
The evolution of the main timing parameters is plotted in Fig. \ref{fig:timing} ({\it left}). 
The QPO centroid frequency increased as a function of time until MJD 56210, 
then it decreased again to become non detectable after MJD 56215.  At the same time, the noise break frequency increased 
and the total fractional rms decreased.  
On the other hand, the QPO fractional rms increased until  roughly MJD 56188 and then it decreased. 
This is consistent with a typical evolution of HS and then HIMS. 

In the harder {\it{RXTE}}/PCA energy band (2--40 keV), \citet{munoz11a} found that below 5\% total fractional rms GX 339-4 entered the SIMS. 
Here, no observation shows a value below 5\%. 
Fig. \ref{fig:timing} ({\it right}) shows the QPO centroid frequency as a function of total fractional rms, 
a good indicator to establish the QPO type \citep{motta11}. 
Comparing with Fig. 4 of \citet{motta11}, it is clear that all QPOs observed here are of type C, 
indicating that at least until MJD 56214 the source was in the HIMS (QPO frequency and rms only for those observations where the first QPO was significant have been plotted) .

\begin{figure}
\centering
\includegraphics[height=6cm]{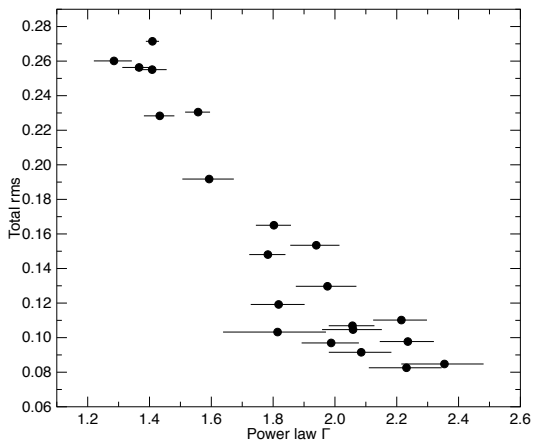}
\caption{Total fractional rms vs power-law spectral slope.
 \label{fig:correl}}
\end{figure}

In order to relate the timing evolution with the spectral properties, 
we show the relation between the \swift/XRT power law spectral index $\Gamma$ and the total fractional rms  (Fig. \ref{fig:correl}). 
The relation between these two parameters is, although noisy, roughly monotonic, without large outliers. 
Similar correlations have been reported in other sources both with spectral fit based parameters such as $\Gamma$ \citep{grinberg14} 
or empirical measures for the spectral shape, such as hardness \citep{munoz11b}. 
The tight correlation between the total fractional rms and the QPO frequency (Fig. \ref{fig:timing}, {\it right})
implies a similar correlation between the QPO frequency and spectral shape.
Overall, the timing parameters seem to trace the spectral behaviour, as expected in the canonical view of state evolution of BHBs.

\begin{figure}
\centering
\includegraphics[height=16cm,width=8cm]{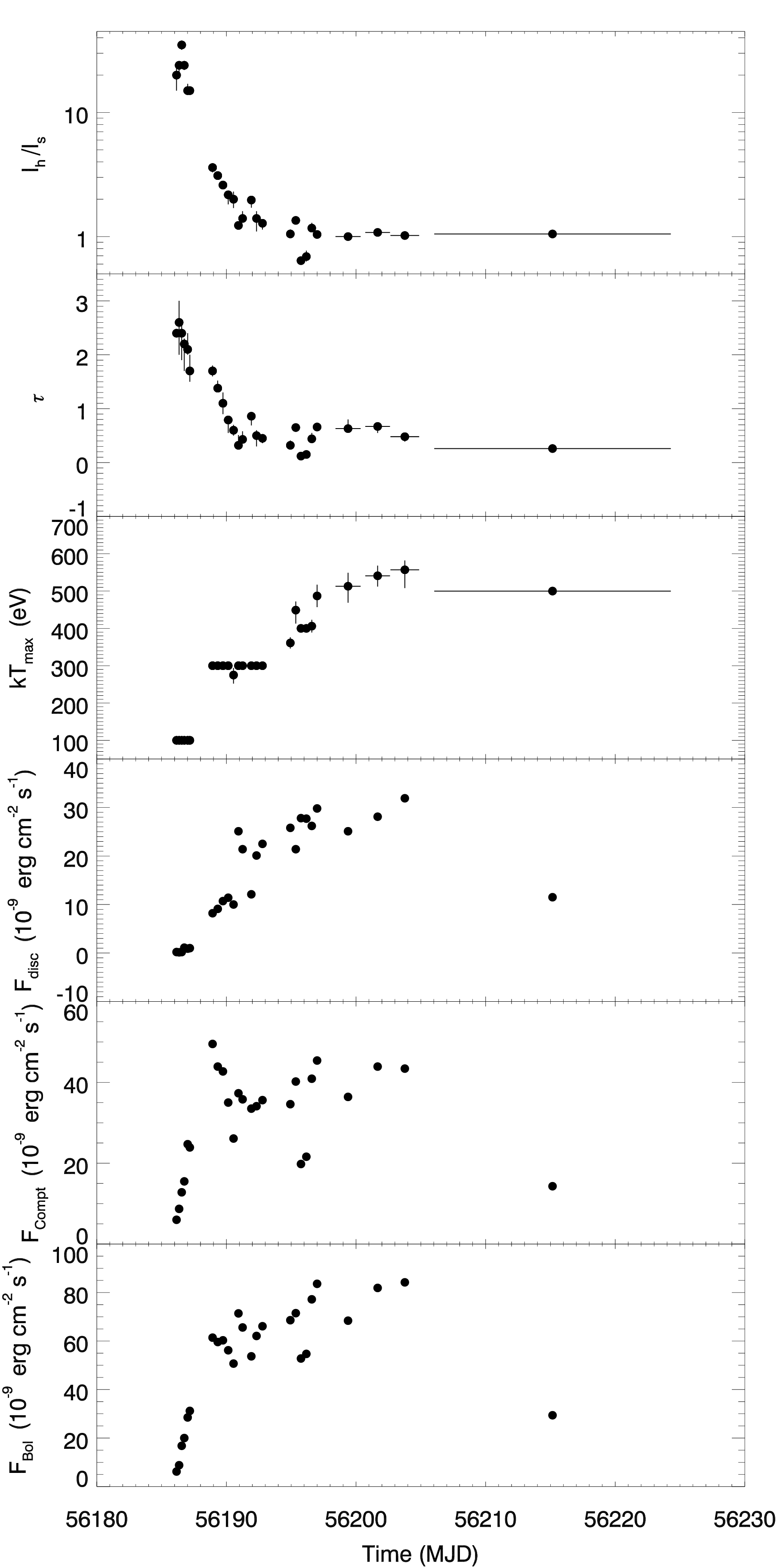}
\caption{Spectral parameters and fluxes evolution obtained with the {\it{eqpair}} model (values reported in Tab. \ref{tab:fit}).
 \label{fig:eqpair_time}}
\end{figure}

\section{Describing broad-band emission with hybrid Comptonisation models}
\label{sec:eqpair}
The IBIS spectra  have been combined (when possible) with the XRT quasi-simultaneous spectra and  fit with {\it eqpair},
the hybrid thermal/non-thermal Comptonisation model developed by \citet{coppi99}.
In {\it eqpair}, the emission of the disc/corona system is modelled by a
spherical hot plasma cloud with continuous acceleration of electrons illuminated by soft
photons emitted by the geometrically thin accretion disc. 
At high energy the distribution of Comptonising electrons is non-thermal,
but at low energies a  thermal population (Maxwellian) is established.
The non-dimensional compactness parameter determines the properties of the plasma: 
\begin{equation}
\label{eq:comp}
\ell = \frac{ \sigma_{T}} { m_{e} c^{3}} \frac{L}{R}
\end{equation}
where  $L$ is a power (luminosity) of the source supplied by different components, $R$ is the radius of the 
sphere, $\sigma_{T}$ is the Thomson cross-section,  $m_e$ is the electron mass and $c$ is the speed of light.
The {\it eqpair} compactness parameters are: $\ell_{\rm s}$, $\ell_{th}$, $\ell_{\rm nth}$ and $\ell_{\rm h}=\ell_{th}+\ell_{\rm nth}$,
corresponding to the power in soft disc photons entering the plasma, thermal electron heating, 
electron acceleration and total power supplied to the plasma, respectively.
The spectral shape strongly depends  on the compactness ratios $\ell_{\rm h}/\ell_{\rm s}$ and $\ell_{\rm nth}/\ell_{\rm h}$ and not on the single parameters.
It is customary to fix $\ell_{\rm s}$ (parameter $\ell_{\rm bb}$ in {\sc{xspec}}) to some reasonable value when fitting data with  {\it eqpair} (e. g. \citealt{gierli99, delsanto08}).
Indeed, as suggested by the developer of {\it eqpair}\footnote{see https://heasarc.gsfc.nasa.gov/xanadu/xspec/models/eqpap4.ps}, 
for Galactic BHBs the best recourse may be simply to leave the soft photon compactness frozen at the unity.
As a cross-check, we fit our spectra with $\ell_{\rm bb}$=10 obtaining no variation in the spectral parameters.
We definitely assumed $\ell_{\rm bb}$=1 (see also \citealt{delsanto13, joinet07, malzac06})
which implies that the variations of the ratio $\ell_{\rm h}/\ell_{\rm s}$  is only due to changes in $\ell_{\rm h}$.  
However, we do not know if this is what really happens  or if it is the $\ell_{\rm s}$ which changes (there is no possibility to fix $\ell_{\rm h}$ in {\sc{xspec}}).

High values of $\ell_{\rm h}/\ell_{\rm s}$ (greater than 10) and $\tau$ ($> 1$) are usually measured in spectra of BH binaries in hard state (e. g. \citealt{ibragimov05, delsanto13}).
Thus, spectra between A1 and A6 of \sou\ (Tab. \ref{tab:fit}, Fig. \ref{fig:eqpair_time}) show parameters typical of the HS, as also indicated by the timing features
 (see energy spectrum and PDS in Fig. \ref{fig:spec_time}, upper panel),
 while starting from group B1 to D4 typical parameters of the HIMS, i.e. $\ell_{\rm h}/\ell_{\rm s}$ of the order of the unity, are inferred (Tab. \ref{tab:fit}).
Interestingly, the Thomson optical depth of the corona is continuously decreasing as the outburst evolves, 
dropping from values of about 2.5 at the beginning of the HS down to 0.1 in the HIMS.
It is worth noting that this parameter is also very variable within the HIMS itself, possibly associated to the variation of the size of the emitting region (see Sec. \ref{sec:discussion}).

Usually, the reflection component  (parameter $\Omega$/$2\pi$) is faint \citep{zdz99} or sometimes absent in hard state.
Such a signature appears stronger in the intermediate states and becomes very strong in soft states.
This is expected when the system evolves from a geometry where the reflecting disc 
is truncated at a large distance from the black hole to a situation where the accretion disc is sandwiched by the illuminating corona \citep{done07}. 
In the latest version of {\it eqpair} (v. 1.10), the code uses {\it ireflct},  a convolution model for reflection from ionised material according to \citet{mz95},
and {\it rdblur} for rotational blurring.\\
In \sou, this component appears in  spectrum A3. Introducing the reflection component in the model to our spectra improves the fit significantly
(F-test probability= 2.3 $\times 10^{-8}$).
When the statistical quality of the spectrum was too low to constrain the reflection parameter, this was fixed at $\Omega$/$2\pi=1$ (see Tab. \ref{tab:fit}).

As a disc model (internal to {\it eqpair}), we use an extension of the {\it diskbb}, 
i.e. a pseudo-newtonian disc ({\it diskpn} in {\sc xspec}) including corrections for temperature distribution near the black hole.
The seed photon temperature kT$_{\rm max}$ was frozen at 100 eV in the harder spectra (A1--A6).
Then, from B1 to B4 and from B6 to B10, kT$_{\rm max}$ was fixed at 300 eV which is comparable with the value inferred from the B5-008 spectrum.
According to the temperature obtained for spectra C1-011 and C2-012, we fixed  kT$_{\rm max}$ at 400 eV in C3 and C4.
Finally, from group C5-013 onwards, the disc black-body temperature increases up to $557^{+25}_{-49}$ eV.

The non-thermal electrons are injected with a power-law distribution $\gamma^{-G_{inj}}$, 
with Lorentz factors ranging from $\gamma_{min} = 1.3$ to $\gamma_{max} = 1000$.
Because of the low statistics at high energy, it is not possible to give constraints to the $G_{inj}$ parameter,
so as expected from shock acceleration models, we  fixed  $G_{inj}$  at 2.5.

Concerning the fraction of Comptonization by non-thermal electrons, we found an indication of non-negligible values of the parameter $l_{\rm nth}/l_{\rm h}$ 
only in the spectra D1-016, D2-019 and D3-020. The fraction of  electron acceleration to the total power supplied to the plasma is about 0.25 (see Sec. \ref{sec:high} for an extensive discussion).
 
We have also estimated the bolometric fluxes (0.1-1000 keV) for each sub-group, and the fluxes of the two main spectral
components over the whole band, such as the geometrically thin disc black-body and the Comptonisation by the hybrid corona electrons (see Tab. \ref{tab:fit}).

\subsection{Focussing on the non-thermal Comptonisation component}
\label{sec:high}
 Despite the statistically acceptable fits obtained when fixing $l_{\rm nth}/l_{\rm h}$ at 0 in spectra D1-016, D2-019 and D3-020, 
we have noted some residuals in the fits. 
In order to investigate further the possible presence of the non-thermal component during the HIMS, 
the spectrum D1-016 has been fit with the {\it compPS} model (assuming a spherical geometry) with Maxwellian electron distribution \citep{pou96}.
This resulted in a reduced $\chi^{2}$ of 1.07 (652 d.o.f.). Adding a power-law component with a slope of $2.5 \pm 0.1$, the reduced
$\chi^{2}$ definitely improved down to 0.95 (650 d.o.f.). The F-test probability that this improvement was by chance is $8.9 \times 10^{-18}$.
In addition, assuming a hybrid electron distribution in {\it compPS}, we obtained a $\chi^{2}_{\rm red}$=0.99(651).
However, because of the low statistics at high energy, we did not manage to constrain the  electron power-law index ($\Gamma_{e} < 5$).
Similar results have been obtained for the two spectra D2-019 and D3-020.

\begin{figure}
\centering
\includegraphics[height=9cm, width=5cm, angle=+90]{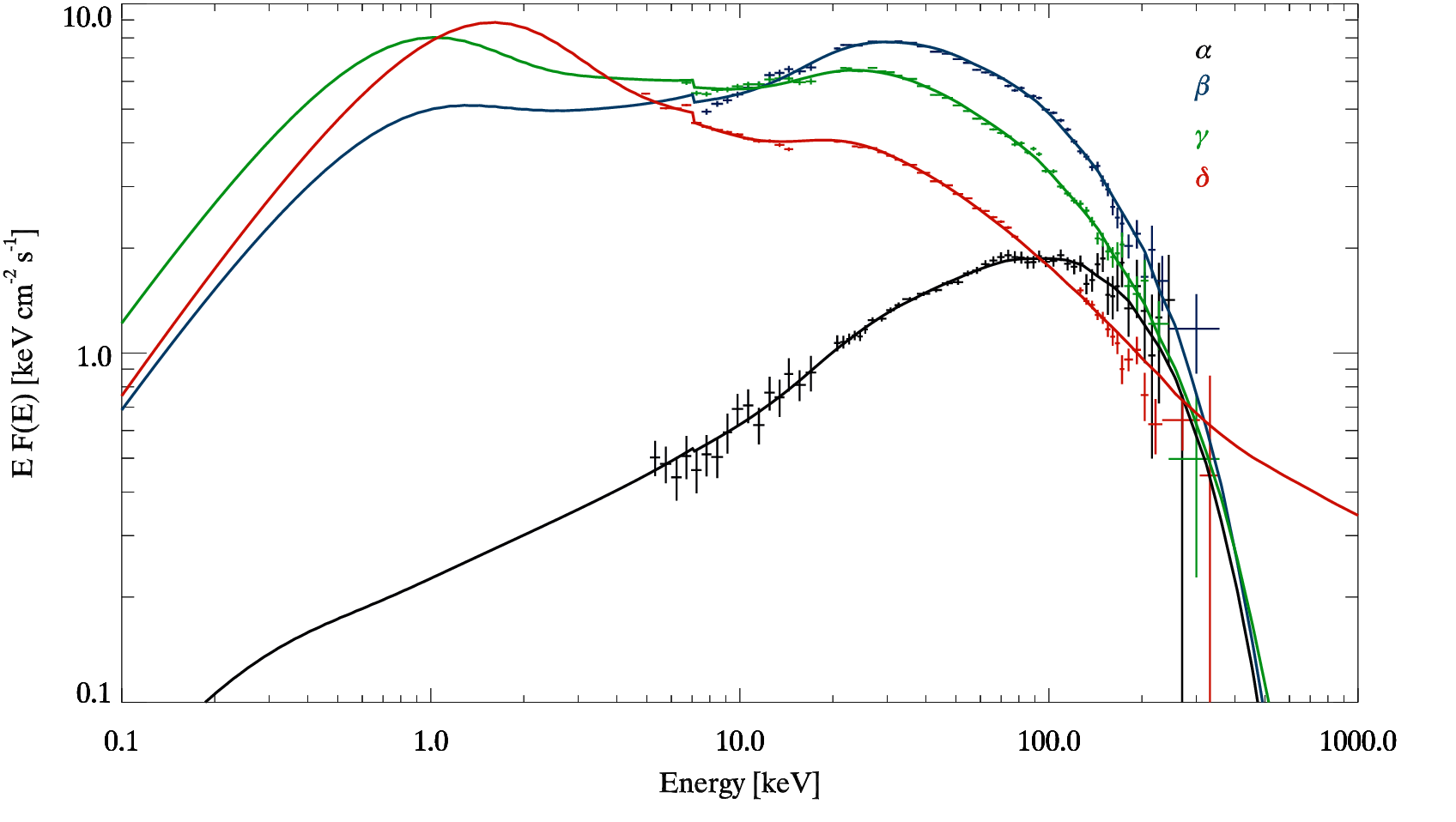}
\caption{Quasi-simultaneous IBIS and JEM-X energy spectra averaged on the periods $\alpha$ (black), $\beta$ (blue), $\gamma$ (green), $\delta$ (red) 
fitted with {\it{eqpair}} (For the interpretation of the references to colour in this figure caption, the reader is referred to the electronic version of this paper).
 \label{fig:energy_four}}
\end{figure}

In order to better quantify the Comptonisation fraction by non-thermal electrons,
we averaged the 26 IBIS spectra to increase the statistics at high energy (above 100 keV). 
We identified 4 bigger groups showing similar values of the parameter $\ell_{\rm h}/\ell_{\rm s}$:
 A, B1--B5, B6--C2, C3--D4, namely $\alpha$, $\beta$, $\gamma$, $\delta$ (see Fig. \ref{fig:lc}, bottom).
We combined them with the JEM-X2  (quasi) simultaneous spectra (IBIS cross-normalization factor of about 0.9) 
and performed the fitting procedure with both {\it eqpair} and {\it compPS} (best-fit in Tab. \ref{tab:fit_compPS}).
First, in {\it eqpair} we fixed $\ell_{\rm nth}/\ell_{\rm h} = 0$ for all spectra obtaining an unacceptable reduced $\chi^{2}$ of 4.72(49) by fitting the spectrum $\delta$. 
In the latter, a fraction of electron acceleration to the total power supplied to the plasma of $0.59^{+0.02}_{-0.05}$ is required.
In Fig. \ref{fig:energy_four}, the four JEM-X2--IBIS energy spectra and best-fit models are shown.

Then, we fitted the four JEM-X2/IBIS spectra with {\it compPS} in the case of a Maxwellian electron distribution.
Also in this case, the fit of the last spectrum was unacceptable giving a reduced $\chi^{2}$ of 9.3(49).
So that, assuming a hybrid thermal/non-thermal electron population also in {\it compPS} provides a good fit (see Tab. \ref{tab:fit_compPS}).

\section{Discussion}
\label{sec:discussion}
 During the \sou\ outburst, we have observed only the HS-to-HIMS transition
which occurred most likely between MJD 56187.2 and MJD 56188.7 (in the gap between groups A and B, see Fig. \ref{fig:lc})
when the QPO rms started to decrease (see Fig. \ref{fig:timing}, left). We did not witness the SIMS, which is characterised by 
noise that is even weaker than we observe, i.e. below 5\%.
Even though the decreasing X-ray flux seems to indicate an evolution back to the hard state (see \citealt{kalemci14}),
we cannot exclude that a secondary maximum (i.e. a transition to the soft state) occurred during the gap in  \integral\ and \swift\ observations (2012 November-2013 January).
This second peak in the X-ray light curve occurring several months after the first one has been observed in other transient BH binaries \citep{chen93, castro97}.
However, radio band observations strengthen the failed outburst hypothesis, since they indicate that the self-absorbed radio jets were never fully quenched \citep{curran14}.

 A correlation between the timing and spectral parameters (i.e. total rms and power-law slope) obtained with the XRT data  has been derived.
The total fractional rms diminishes during a transition from hard to softer states for two reasons. 
The first reason is an increase of the amount of soft photons coming from an increasingly bright accretion disc in the energy band where 
the rms is measured (0.3-10 keV here). The disc photons are generally speaking not variable (but see \citealt{wilkinson09}) 
and therefore they dilute the variability carried by the harder photons, causing a general decrease of the fractional rms. 
The second reason is an intrinsic decrease of the variability of the hard photons. In the context of the truncated disc 
model and rigid precession of the inner flow (see e.g. \citealt{ingram09}, \citealt{motta15}), 
both the noise and the QPOs observed in the HS and in the HIMS are produced in  the inner part of the accretion flow that precesses in a rigid fashion following the Lense-Thirring effect. 
The QPOs are produced via the modulation imprinted by the precession to the emission, while the noise, which carries most of the variability, 
is produced via propagation of mass accretion rate fluctuations in the same flow, fundamentally causing intrinsic brightness variability. 
The decrease of the outer radius of the inner flow (linked to the truncation radius of the disk) causes the loss of the variability at 
low frequencies as the source moves to softer states. This effect combines with a cooling of the inner hot flow, 
that is responsible for the decrease in the number of hard photons (both producing the noise and the QPOs), causing a further decrease of the total fractional variability.

Fitting the XRT/ISGRI spectra with a hybrid Comptonisation model ({\it{eqpair}}), we have observed a decreasing of the $\ell_{\rm h}/\ell_{\rm s}$ parameter during the outburst.
We do not know whether the $\ell_{\rm h}/\ell_{\rm s}$ variation is due to changes in the heating rate 
of the corona (related to $\ell_{\rm h}$), changes in the luminosity of the disc ($\ell_{\rm s}$), or both.\\
In \sou, we observe a change of the coronal optical depth by a factor of 20 which could be associated either with a change in electron density 
 or in the size of the corona (since $\tau\propto n_{e}R$). 
A constant electron density ($n_{e}$) would imply that the corona is larger in the HS than in the HIMS by up to a factor of $\sim$20.

Looking at the flux variation (Fig. \ref{fig:eqpair_time} and Tab. \ref{tab:fit}), it is clear that the bolometric flux increases in a few days and is mainly driven by the Comptonisation component.
Then, during the transition to the HIMS, the  Comptonisation luminosity, which is related both to the electron heating and electron acceleration, varies by only a factor of 2,
while the disc flux increases by a factor of 10 or even more as the outburst proceeds.
This would imply that the HS-to-HIMS spectral transition is driven by changes in the soft photon flux in the corona 
also associated with the increase of disc temperatures (which is also observed). The increasing of the soft photon flux 
should be also responsible of the cooling of the corona which reflects in the hard X-ray spectral softening observed \citep{gilfanov99}.
In contrast, the heating rate of the electrons in the corona appears to change by only a factor of 2. \\
Although other models  (i.e. dynamic accretion disc corona) cannot be ruled out,
these results are consistent with the so-called truncated disc model  \citep{done07}, 
also supported by the increase of the Compton reflection and by the larger QPO frequencies in the HIMS compared to the HS \citep{zdz99, gilfanov99}.
  
 We did not find in the HS any contribution from the non-thermal electron Comptonisation  as observed by \integral\ in a number of BHBs (e.g. \citealt{bouchet09, pott08, droulans10}).
This is possibly due to the lack of long observations in a stable spectral state, which results in low statistics carried out above 200 keV in the IBIS spectra.
However, as expected, this component is observed in the HIMS, when the thermal electrons of the corona cool and the non-thermal Comptonisation occurs.

\section{Conclusions}
We have presented spectral and timing analysis of \integral\ and \swift\ observations of the transient BHB \sou.
Data covered the first part of the outburst spanning from 2012 September 16 until October 30.
Our results can be briefly summarised:
\begin{enumerate}
\item[1)] Despite the source brightness (up to L$_{\rm bol} \sim 0.4$L$_{\rm Edd}$), \sou\ never reached the SS state,
              increasing the number of "failed outbursts" observed in BHBs.
              This implies that the low luminosities observed in most of the previous failed outburst is not the only discriminant for such a phenomenon.
 \item[2)]  The truncated disc model is supported by the evolution of the spectral parameters as the source goes from the HS to the HIMS and
                  by the high flux of disc photons combined with the increasing of the black-body temperature.
                 In addition, the timing results and, in particular, the correlation between the spectral softening and the total fractional rms
                  strengthen this scenario.

\item[3)] We did not find any contribution from non-thermal Comptonisation in the HS. On the contrary, a non-thermal fraction of 0.6 
              contributes to the total Comptonisation emission in the HIMS,
              when the cooling of thermal electrons is also observed.
\end{enumerate}

\section*{Acknowledgments}
MDS thanks Milvia Capalbi and Carlo Ferrigno for useful discussion on \swift/XRT and \integral\ data analysis, respectively.
TMB acknowledges support from INAF PRIN 2012-6.
JAT acknowledges partial support from the {\em Swift} Guest Observer program
through NASA grants NNX13AJ81G and NNX14AC56G.
VG acknowledges funding support by NASA through the Smithsonian Astrophysical Observatory (SAO) contract SV3-73016 to MIT
for Support of the Chandra X-Ray Center (CXC) and Science Instruments; CXC is operated by SAO for and on behalf of NASA under contract NAS8-03060.
JR  acknowledges funding support from the French Research National Agency: CHAOS project ANR-12-BS05-0009 (\texttt{http://www.chaos-project.fr}), 
and from the UnivEarthS Labex program of Sorbonne Paris Cit\'e (ANR-10-LABX-0023 and ANR-11-IDEX-0005-02).
TMD acknowledges support by the Spanish Ministerio de Economia y competitividad (MINECO) under grant AYA2013-42627.

\begin{table*}
\begin{center}
\caption{Best-fit parameters of the joint IBIS/ISGRI and XRT spectra when available (on the top) and of the joint IBIS and JEM-X2 spectra (on the bottom). 
Fits have performed with {\sc eqpair}. 
See text for the parameters description. Values in parenthesis denote parameters fixed during the fits.
Bolometric flux (0.1--1000 keV) as well as fluxes of the two main components, i.e. disc and Comptonisation, have been estimated.
 }\label{tab:fit}
\renewcommand{\arraystretch}{1.3}
\begin{tabular}{lcccccccccc} 
\hline
\hline
Group & N$_{\rm H}$ & $l_{\rm h}/l_{\rm s}$ & $\tau_{\rm p}$& $\Omega$/${2\pi}$&$l_{\rm nth}/l_{\rm h}$ &k$T_{\rm max}$ & $\chi^2_{\nu}$(dof) &\multicolumn{3}{c}{ $F  \times 10^{-9}$} \\	
 IBIS-XRT & $\times 10^{22}$ cm$^{-2}$    &                           &                    &      &      &          [eV]         & &  \multicolumn{3}{c}{[\ergcms]} \\
      &             &              &                    &      &      &                            &                                     &          Bol & disc & Compt \\
\hline

A1 & - &  $20^{+3}_{-5}$ &  $< 2.4$ &(0) &  (0) & (100)  & 1.10(39) & 6.2  & 0.2   &  6.0     \\
A2 & - &  $24^{+2}_{-5}$ &  $  2.6^{+0.4}_{-0.6}    $ &(0) &(0) & (100) &  0.80(46) & 8.8 &  0.14  & 8.7  \\
A3-000 & $1.36\pm 0.02  $   &   $35 \pm 3$ &  $2.4^{+0.2}_{-0.5}$ & $0.7 \pm 0.2$ &(0) & (100) &  1.05(730)&16.8 & 0.19 &  12.8  \\
A4-002 &$1.35^{+0.03}_{-0.02}$&  $24\pm 2$ & $2.2^{+0.1}_{-0.5}$&  $0.5^{+0.2}_{-0.1} $ &(0)   &  (100)  & 0.85(154)&  20.0 &  1.1  &  15.5\\
A5 & - & $15^{+2}_{-1}   $ &  $ 2.1^{+0.3}_{-0.1} $   &   $<0.4$ & (0) &  (100)  & 0.98(45)&   28.5 &  0.9  &  24.7\\
A6&  -  &  $15.0^{+0.6}_{-0.7}$ & $1.7^{+0.3}_{-0.2}$&  $0.7 \pm 0.3  $&(0) &   (100)  & 0.77(47)&31.2  &  1.0  &  23.9  \\

B1& - & $3.6^{+0.2}_{-0.3}$  & $1.7 \pm 0.1$&  $0.2^{+0.2}_{-0.1}  $&(0) &  (300)  & 0.66(45)&   61.4 & 8.2  &  49.5 \\
B2& - &  $3.1 \pm 0.1$ & $1.38^{+0.14}_{-0.07}$&  $0.54 ^{+0.09}_{-0.24}  $& (0)   &  (300) & 0.97(45)& 59.6  & 9.1  & 43.9 \\
B3& - & $2.6 \pm 0.2$  & $1.1^{+0.2}_{-0.3}$&  $0.6 ^{+0.5}_{-0.1} $&(0)   &(300)   & 0.71(45)&  60.3 & 10.7  & 42.7\\
B4& - & $2.17^{+0.03}_{-0.35}$  & $0.79^{+0.06}_{-0.24}$&  $1.07^{+0.16}_{-0.08} $& (0)   & (300)   & 0.81(48)&  56.2 & 11.4  & 35.0 \\
B5-008 &$1.47\pm 0.05$ & $2.0 \pm 0.3$  & $0.6 \pm 0.1 $&  $ < 1.7  $& (0)    & $275^{+19}_{-23}$    & 1.04(709)& 50.7 & 10.0  &  26.1 \\
B6& - & $1.23^{+0.05}_{-0.03}$  & $0.32^{+0.19}_{-0.03}$&  $(1)$&$(0)$    & (300) &  1.11(46)   & 71.4 & 25.1 &  37.3\\
B7& - & $1.4^{+0.2}_{-0.1}$  &  $0.43^{+0.15}_{-0.07}$&  $ (1)$& $(0)$ & (300)   & 0.69(36)&  65.6 &  21.4 &   35.8 \\
B8& - & $1.97^{+0.07}_{-0.26}$  & $0.86^{+0.05}_{-0.17}$&  (1) & $ (0) $ & (300)  &   0.83(48)& 53.7 &  12.1 &   33.5  \\
B9& - & $1.4^{+0.2}_{-0.3}$  &$0.5^{+0.1}_{-0.2}$&  (1)& $ (0) $ &(300) &  0.80(46)&  62.1 &  20.1&    34.1 \\
B10& - & $1.28^{+0.04}_{-0.14}$  &$0.45^{+0.02}_{-0.09}$&  $(1)$& $ (0) $ & (300) &   0.79(46)& 66.1&   22.5 &   35.6 \\

C1-011 &  $1.88^{+0.04}_{-0.07}$ & $1.05^{+0.09}_{-0.07}$  & $ 0.32^{+0.09}_{-0.07}$&  (1) &$ (0) $    & $ 361 \pm 15 $   &  1.03(659)&   68.6 & 25.8  & 34.6 \\
C2-012 &   $1.44^{+0.04}_{-0.02}$&  $1.35\pm 0.06$  & $ 0.65^{+0.02}_{-0.03}$&  $(1)$&$ (0) $    & $ 449^{+23}_{-36} $   & 1.04(695)&  71.5 &  21.4  & 40.2 \\
C3& - & $0.64 \pm 0.04 $  & $0.12^{+0.03}_{-0.05}  $&  $ (1)$&$ (0) $    &  (400)    &  1.08(46)& 52.8  & 27.8   & 19.8   \\
C4& - & $0.69_{-0.06}^{+0.08}$  & $ 0.15^{+0.07}_{-0.03} $&  $ (1) $&$(0)$    &  (400)    & 0.72(46)& 54.7  & 27.7   & 21.6 \\
C5-013&  $1.66^{+0.01}_{-0.07}$   &$1.17^{+0.12}_{-0.06}$  & $  0.44^{+0.11}_{-0.02}$  &  (1)& (0)    &  $406 \pm 17 $   & 1.03(705)&  77.2   & 26.2   & 40.9 \\
C6-014 &  $1.56^{+0.04}_{-0.05}$ &$1.04^{+0.04}_{-0.05}$  & $ 0.66^{+0.04}_{-0.05} $&  $(1) $&$  (0) $    &  $487 \pm 30 $     & 0.99(703)&  83.6  & 29.8   &  45.4  \\

D1-016 &$1.47^{+0.07}_{-0.02}$ & $1.0^{+0.04}_{-0.05}$  & $  0.63^{+0.17}_{-0.03}  $&$ (1) $  &$ 0.24^{+0.07}_{-0.03} $    &    $ 513^{+36}_{-44} $   &  0.95(651)&  68.4&  25.1 &  36.4\\
D2-019 & $1.45\pm 0.04 $  &$1.08^{+0.04}_{-0.05}$  & $  0.67^{+0.05}_{-0.12}  $ &  $ (1)$  & $ 0.24^{+0.02}_{-0.04}$    &    $ 541^{+27}_{-29}   $   &  0.93(656)&  81.9 & 28.1  & 43.9 \\
D3-020 &$1.44\pm 0.04$ & $1.02^{+0.05}_{-0.02}$  & $  0.48^{+0.06}_{-0.09}  $ &  $ (1)$  &  $ 0.23^{+0.03}_{-0.10} $ &  $557^{+25}_{-49} $   & 0.97(677)&   84.2 & 31.9 &  43.4\\
D4 &- & $1.05 ^{+0.02}_{-0.05}$  & $  0.26^{+0.02}_{-0.05}  $ &  $ (1)$  & $(0) $  &  $(500) $   & 0.74(48)&   29.4  & 11.5   &14.3 \\

\hline
\hline
\end{tabular}
\vspace*{3 cm} 
\end{center}
\end{table*}

\begin{table*}
\begin{center}
\caption{Best-fit parameters of the joint IBIS and JEM-X2 spectra. 
Fits have performed with {\sc eqpair} (first line) and {\sc compPS} (second line). In the latter, seed photons are multicolor disk black-body (kT$_{bb}$), 
$\tau$-y is the Compton parameter and $\Gamma_{e}$ is the electron power-law index.
Values in parenthesis denote parameters fixed during the fits. }\label{tab:fit_compPS}
\renewcommand{\arraystretch}{1.3}
\begin{tabular}{lccccccc} 
\hline
\hline
Group & $l_{\rm h}/l_{\rm s}$ & $\tau_{\rm p}$  & $\Omega$/${2\pi}$ & $l_{\rm nth}/l_{\rm h}$ &k$T_{\rm max}$                  & k$T_{e}$ &    $\chi^2_{\nu}$(dof)  \\	
          &                                   &       $\tau$-y      &                            &        $\Gamma_{e}$         &        k$T_{\rm bb}$             &                  &                        \\
          &                                   &                                       &      &      &          [eV]   &         keV                &                    \\
\hline
$\alpha$& $22\pm 2$  & $ 2.2^{+0.1}_{-0.4}$&  $ 0.3^{+0.3}_{-0.1} $&  (0)    &  (100) &      -   & 0.93(60)   \\
              &       -            &           $<3$              &    $0.6 \pm 0.2$          &     -   &    (100)  &    $65^{+5}_{-2}$       & 1.01 (60)  \\

$\beta$ & $2.7\pm 0.1$  & $ 1.12 \pm 0.05$&  $  0.8 \pm 0.1$&  (0)     &  (300) &   -        &    1.03(57)    \\
     &       -            &           $1.8\pm 0.2$             &    $0.9 \pm 0.2$          &     -   &    (300)  &    $57^{+7}_{-5}$      & 0.98 (57)  \\

$\gamma$  & $1.57^{+0.06}_{-0.05}$    & $ 0.69 \pm 0.02 $&  $  0.7 \pm 0.1$&  (0)     &  (300) &   -        &    0.91(59)    \\
     &       -            &           $0.9^{+0.2}_{-0.3}$             &    $0.9 \pm 0.2$          &     -   &    (300)  &    $82^{+23}_{-13}$      & 0.91(59)  \\

$\delta$ & $0.75^{+0.13}_{-0.06}$    & $ 0.68^{+0.08}_{-0.16} $&  (1)  &  $0.59^{+0.02}_{-0.05}$    &  (500) &   -        &    0.95(48)    \\
     &       -            &           $1.4^{+0.7}_{-0.4}$             &    (1)          &     $4^{+0.2}_{-0.3}$  &    (500)  &    $<26$      & 0.75(48)  \\

\hline
\hline\end{tabular}
\vspace*{3 cm} 
\end{center}
\end{table*}

\end{document}